\documentclass{jfm}
\usepackage{graphicx}
\usepackage{psfrag, mathrsfs, subfig, amsmath, amssymb, natbib, multirow, bbding}

\DeclareFontFamily{OT1}{pzc}{}
\DeclareFontShape{OT1}{pzc}{m}{it}{<-> s * [1.10] pzcmi7t}{}
\DeclareMathAlphabet{\mathpzc}{OT1}{pzc}{m}{it}

\newcommand\pd[2]{\frac{\partial#1}{\partial#2}}
\newcommand\vc[1]{\boldsymbol{#1}}

\newcommand\ud{\textrm{d}}
\newcommand\Ro{\textrm{Ro}}

\makeatletter
\newcommand\Rmnum[1]{\expandafter\@slowromancap\romannumeral #1@}
\makeatother

\title[Hill's Spherical Vortex in a Rotating Fluid]
  {Hill's Spherical Vortex in a Rotating Fluid}

\author[M.~M.~Scase \& H.~L.~Terry]
{M.\ns M.\ns S\ls C\ls A\ls S\ls E \& H.\ns L.\ns T\ls E\ls R\ls R\ls Y}

\affiliation{School of Mathematical Sciences, University of Nottingham, Nottingham NG7 2RD}

\date{\today} 

\begin{document}

\maketitle

\begin{abstract}
A popular model for a generic fat-cored vortex ring or eddy is Hill's spherical vortex (\textit{Phil.~Trans.~Roy.~Soc.~}A vol. 185, 1894, p.~213).  Here we find an exact solution for such a spherical vortex steadily propagating along the axis of a rotating ideal fluid.  The spherical vortex swirls in such a way as to exactly cancel out the background rotation of the system.  The flow external to the spherical vortex exhibits fully nonlinear inertial wave motion.  We show that above a critical rotation rate, closed streamlines form in this outer fluid region and hence fluid is carried along with the spherical vortex.  As the rotation rate is further increased, further concentric `sibling' vortex rings are formed, counter rotating with respect to the original spherical vortex.
\end{abstract}

%****************************************************************************************

\section{\label{sec:int}Introduction}

%****************************************************************************************

In 1894 Hill published his famous solution for the steady flow of a spherical vortex in an ideal fluid \citep{hill}.  The solution consists of an inner rotational spherical region of fluid that matches onto an outer irrotational region of fluid that extends to infinity.  It was shown that the outer flow was equal to that of steady inviscid flow past a sphere.  Hill's solution was later shown to be the end member of a family of steadily propagating vortex rings \citep{norbury} of varying core thickness that includes `thin-cored' rings \citep[see][]{fraenkel70, fraenkel72} where the rotational fluid is confined within a narrow region that does not extend to the axis.  These solutions, in particular Hill's, have been the focus of a number of stability analyses \citep[see e.g.,][]{moffattMoore, pozrikidis, protasElcrat} that show that in time fluid may be detrained or entrained into the vortex according as to whether it has a prolate or oblate deformation respectively.

As part of his body of work on rotating fluids in the early twentieth century, \citet{taylor22} investigated the response of a rotating fluid to a sphere being steadily towed along its axis of rotation.  Taylor found an exact solution to the Euler equations that supported fully nonlinear inertial waves created by a sphere that translated steadily along the axis of rotation of the fluid but did not precess about this axis in the laboratory frame.  His experiments showed that while the tank rotated but the sphere was not towed, the sphere rotated with the fluid in solid body rotation.  Yet, when the sphere was towed along the axis of rotation it ceased precessing and had no azimuthal velocity in the laboratory frame of reference.  In the analysis of his solution Taylor found that in the limit of the radius of the sphere tending to zero, a structure that resembled Hill's spherical vortex could be observed in the flow, though he described this analogy between his flow and Hill's spherical vortex as `only superficial'.

In the present work we combine the approach of \citet{taylor22} (summarized briefly in \S\,\ref{sec:taylor}) with the solution of \citet{hill} to find an exact solution of the rotating Euler equations with Hill's spherical vortex propagating steadily along the axis of rotation.  Like Taylor's sphere, the spherical vortex does not precess in the laboratory frame, as we show in \S\,\ref{sec:hills}.  In \S\,\ref{sec:results} we show that as the rotation rate of the system is increased above critical rotation rates a series of thin-cored vortex rings form in the outer fluid that counter-rotate with respect to the spherical vortex on the axis of rotation.  In \S\,\ref{sec:conc} we draw our conclusions. 

%****************************************************************************************

\section{\label{sec:modelling}Hill's Spherical Vortex in a Rotating Fluid}

%****************************************************************************************

We consider an inviscid, incompressible fluid with pressure field $p$, velocity field $\vc{u}$ and constant density $\rho$ in a frame of reference that is rotating with rotation vector $\Omega\,\hat{\vc{z}}$, where $\hat{\vc{z}}$ is a unit vector.  The flow is described in terms of spherical polar coordinates $(\sigma, \theta, \phi)$ where $\sigma\geqslant0$ is the radial distance from the origin, $\theta\in[0, \pi]$ is the polar angle and $\phi\in[0, 2\pi)$ is the azimuthal angle.  The spherical coordinate system is aligned such that $\theta = 0$ is in the $\hat{\vc{z}}$-direction.  We seek to model a spherical vortex of radius $a$ and propagation velocity $U\hat{\vc{z}}$, hence we nondimensionalize position as $\vc{x} = a\tilde{\vc{x}}$, velocity as $\vc{u} = U\tilde{\vc{u}}$, and pressure as $p = U^2\rho \,\tilde{p}$, where tildes indicate nondimensional quantities.  Dropping the tildes immediately the nondimensional steady rotating Euler equations that govern the motion are
\begin{equation}  \label{eq:gov}
\nabla\cdot\vc{u} = 0,
\qquad
\left(\vc{u}\cdot\nabla\right)\vc{u} = -\nabla p + \frac{r}{4\Ro^2} \hat{\vc{r}}- \frac{1}{\Ro}\hat{\vc{z}}\times\vc{u},
\qquad
\Ro = \frac{U}{2a\Omega},
\end{equation}
where: we have defined a Rossby number, $\Ro$; $\hat{\vc{r}} = \sin\theta\,\hat{\vc{\sigma}} + \cos\theta\,\hat{\vc{\theta}}$ is a unit vector in the cylindrical radial direction, and $r=\sigma\sin\theta$ is the cylindrical radial position. The non-rotating steady Euler equations are recovered in the limit $\Ro\to\infty$.  

%****************************************************************************************

\subsection{\label{sec:taylor}Taylor's Solution}

%****************************************************************************************

\begin{figure} % Image created with vrPaperFig1.m
\begin{center}
%\setlength{\unitlength}{1pt}
%\begin{picture}(365, 206)(-7, 0)
%\put(0, 0){\includegraphics[bb = 113    79   707   673, height = 200pt, clip = TRUE]{images/taylorPsi.eps}}
%\put(-8, -1){{\scriptsize -2}}	\put(-8, 49){{\scriptsize -1}}	\put(-8, 98){{\scriptsize \phantom{-}0}}	\put(-8, 147){{\scriptsize \phantom{-}1}} 	\put(-8, 198){{\scriptsize \phantom{-}2}}
%\put(-1, -6){{\scriptsize 0}}	\put(48, -6){{\scriptsize 1}}		\put(98, -6){{\scriptsize 2}}				\put(148, -6){{\scriptsize 3}}			\put(198, -6){{\scriptsize 4}}
%\put(-14, 98){$z$}		\put(98, -14){$r$}
%\put(214, 0){\includegraphics[bb = 125    79   574   673, height = 200pt, clip = TRUE]{images/taylorPsi0.eps}}
%\put(213, -6){{\scriptsize 0}}	\put(262, -6){{\scriptsize 1}}		\put(312, -6){{\scriptsize 2}}				\put(362, -6){{\scriptsize 3}}
%\put(287, -14){$r$}
%\put(7, 186){$(a)$}
%\put(221, 186){$(b)$}
%\end{picture}\\[2ex]
%
%vips -o vrPaperFig1.eps -pp3 -O -97pt,-122pt -T 384pt,220pt vrPaper
%\includegraphics[clip = TRUE]{vrPaperFig1.eps}\\[-2ex]
\includegraphics{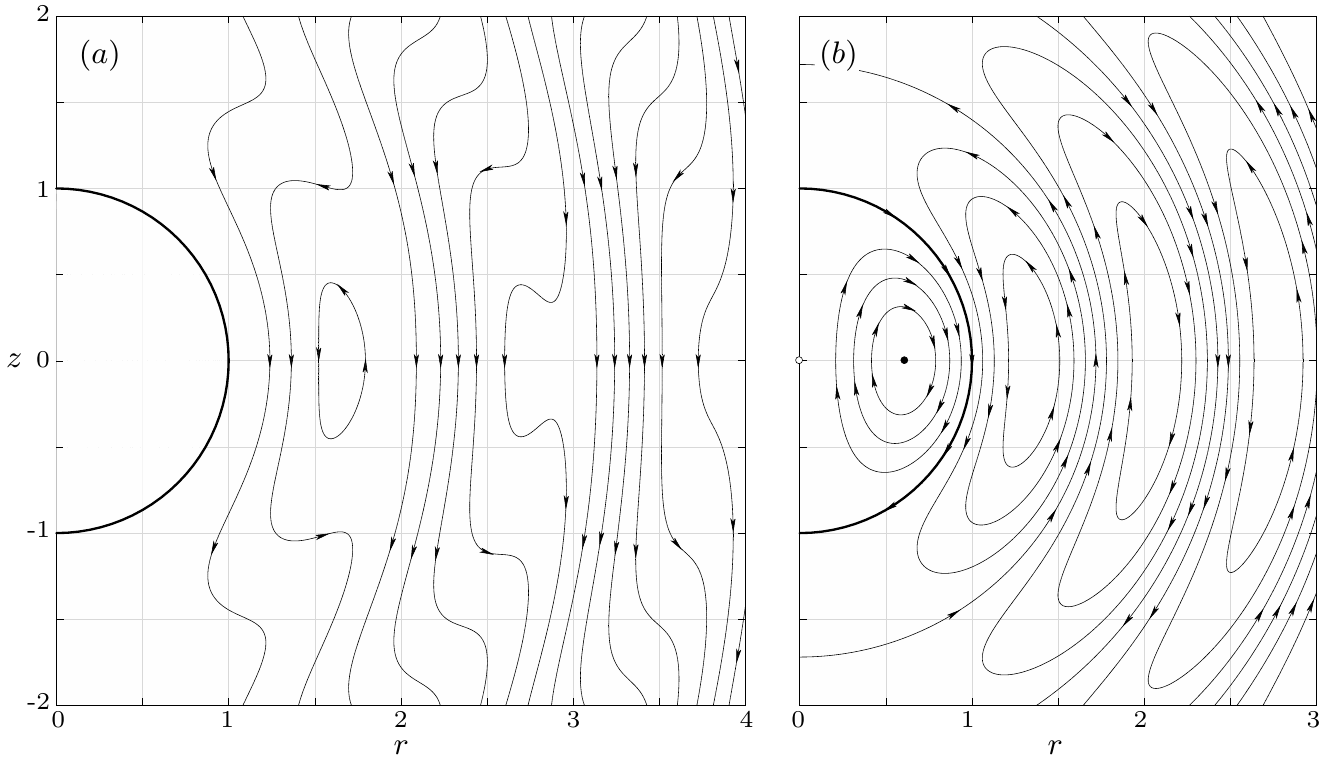}\\[-2ex]
\caption{\label{fig:taylor} $(a)$ Streamlines around a unit sphere translating steadily along the axis of rotation ($r=0$) of a rotating fluid in the reference frame of the sphere.  Closed streamlines show that fluid is being transported with the sphere.  $(b)$ Streamlines due to a sphere of vanishing size (white circle at the origin) translating steadily along this axis of rotation of a rotating fluid in the frame of reference of the main body of fluid.  \citet{taylor22} observed that the streamlines near the origin resemble those of a Hill's spherical vortex \citep{hill} but described the analogy between the flow and a spherical vortex as `only superficial'.}
\end{center}
\end{figure}

\citet{taylor22} considered the axisymmetric flow generated by a non-rotating sphere that is towed with nondimensional velocity $\hat{\vc{z}}$ through a fluid rotating steadily about $\hat{\vc{z}}$.  This may be described by defining an axisymmetric streamfunction, $\psi(\sigma, \theta)$, as in \citet{batchelor}, and writing the fluid velocity in the radial and polar directions respectively as
\begin{subequations} \label{eq:taylorSol}
\begin{equation} \label{eq:taylorSola}
u = \frac{1}{\sigma^2\sin\theta}\pd{\psi}{\theta},
\quad
v = -\frac{1}{\sigma\sin\theta}\pd{\psi}{\sigma}.
\end{equation}
The azimuthal component of the velocity field, $w$, was then posed by \citet{taylor22} to be of the form
\begin{equation} \label{eq:taylorSolb}
w = -\frac{1}{\Ro}\left(\frac{\psi}{\sigma\sin\theta} + \frac{\sigma\sin\theta}{2}\right).
\end{equation}
\end{subequations}
(The second term in brackets is included here as we are working in the non-inertial frame of reference of the rotating fluid, whereas \citet{taylor22} was working in the inertial `laboratory' frame of reference.)  The incompressibility condition (\ref{eq:gov}$a$) is automatically satisfied.  \citet{taylor22} proceeded by posing that the streamfunction be of the separable form $\psi = f(\sigma)\sin^2\theta$.
Substitution into the curl of (\ref{eq:gov}$b$), thus removing the pressure gradient, leads to either the trivial solution $f=0$ or that $f$ must satisfy
\begin{equation} \label{eq:taylor8}
\sigma^3f''' - 2\sigma^2f'' - 2\sigma f' + 8f + \frac{\sigma^2}{\Ro^2}\left(\sigma f' - 2f\right) = 0,
\end{equation}
where a dash denotes differentiation with respect to $\sigma$ \citep[see (8)][]{taylor22}.  In the frame of reference rotating with the fluid, but translating with a sphere of radius $\delta$, the boundary conditions are 
\begin{subequations} \label{eq:taylorbcs}
\begin{equation}
\vc{u}(\delta, \theta) = \left(0,0,-\frac{\delta\sin\theta}{2\Ro}\right),
\qquad
\lim_{\sigma\to\infty} \vc{u}(\sigma, \theta) = \left(-\cos\theta, \sin\theta, 0\right).
\end{equation}
Equivalently, in terms of $f$
\begin{equation} \label{eq:taylorbcsf}
f(\delta) = 0,\quad
f'(\delta) = 0, \quad
\lim_{\sigma\to\infty}\frac{f(\sigma)}{\sigma^2} = -\frac{1}{2}.
\end{equation}
\end{subequations}
The corresponding pressure field is given by
\begin{equation} \label{eq:pressure}
p(\sigma, \theta) = \left(2ff'' - f'^2 + \frac{f^2}{\Ro^2}\right)\frac{\sin^2\theta}{2\sigma^2} - \frac{2f^2}{\sigma^4} + \frac{1}{2},
\end{equation}
where the pressure $p\to(8\Ro^2)^{-1}\sigma^2\sin^2\theta$, the hydrostatic pressure field, as $\sigma\to\infty$.
The solution to \eqref{eq:taylor8} that satisfies \eqref{eq:taylorbcsf} may be written as
\begin{multline}
f(\sigma) = -\frac{\sigma^2}{2} + \frac{1}{2\sigma}\left\{\left[\delta^2\sigma - 3\Ro^2(\sigma-\delta) \right]\cos\left(\frac{\sigma - \delta}{\Ro}\right)
\right. \\ \left.
+\left[3\Ro^2 - \delta^2 + 3\delta\sigma\right]\Ro\sin\left(\frac{\sigma - \delta}{\Ro}\right)\right\},
\end{multline}
and this can be shown to be equal to Taylor's solution.  The streamlines for $\delta = 1$, $\Ro = (2\pi)^{-1}$ are shown in figure \ref{fig:taylor}$a$.  The nonlinear wavefield in the fluid can be observed \citep[cf.~figure 2][]{taylor22} as can the closed streamlines that show that fluid is carried with the sphere.  

\citet{taylor22} observed that the solution still exhibits waves even in the limit that the sphere has a vanishingly small radius, i.e., in the limit $\delta\to0$.  Figure \ref{fig:taylor}$b$ shows this solution, in the laboratory frame of reference.  The sphere is instantaneously located at the origin, indicated by a white circle.  \citet{taylor22} observed that the streamlines near the origin resemble those of a Hill's spherical vortex \citep{hill}.  Here the rotation rate has been chosen such that this apparent spherical vortex has a unit radius (given by the largest $\Ro$ that satisfies $\Ro\sin(\Ro^{-1}) = \cos(\Ro^{-1})$, i.e., $\Ro \approx 0.223$).   The boundary of the apparent spherical vortex is shown in bold.  \citet{taylor22} describes this analogy with Hill's spherical vortex as superficial as the vortex can only exist when the flow is rotating and a vanishingly small sphere is being steadily towed along the axis of rotation. Furthermore, the size of the `spherical vortex' changes according to the rotation rate.  In his experiments \citet{taylor22} noticed that a steadily towed sphere that was free to rotate about the axis of rotation of the fluid in fact did not rotate at all but translates with zero azimuthal velocity in the inertial laboratory frame.  Motivated by this observation we now show that it is possible for a Hill's spherical vortex to exist in a rotating fluid.

%****************************************************************************************

\subsection{\label{sec:hills}Hill's spherical vortex in a rotating fluid}

%****************************************************************************************

The classical Hill's spherical vortex is a spherical region of rotational fluid that propagates through an irrotational ambient fluid.  The solution is found by constructing solutions inside and outside the spherical vortex and enforcing pressure and velocity continuity across the boundary of the two regions.  In a frame of reference moving with the vortex ring, Hill's solution is given by
\begin{subequations} \label{eq:Hsol}
\begin{equation} \label{eq:uH}
\vc{u} = \left\{
\begin{array}{ll}
\displaystyle
\frac{3\cos\theta}{2}(1- \sigma^2) \hat{\vc{\sigma}} - \frac{3\sin\theta}{2}(1 - 2\sigma^2)\hat{\vc{\theta}} &\sigma \leqslant 1 \\[2ex]
\displaystyle
-\frac{\cos\theta}{\sigma^3}(\sigma^3 - 1)\hat{\vc{\sigma}} + \frac{\sin\theta}{2\sigma^3}(2\sigma^3 + 1)\hat{\vc{\theta}} & \sigma > 1.
\end{array}\right.
\end{equation}
\begin{equation} \label{eq:pH}
p =  \left\{
\begin{array}{ll}
\displaystyle
-\frac{9\sigma^2(3 - 2\sigma^2)}{8} \sin^2\theta + \frac{9\sigma^2(2 - \sigma^2)}{8}-\frac{5}{8} & \sigma \leqslant 1 \\[2ex]
\displaystyle
-\frac{3(4\sigma^3 - 1)}{8\sigma^6}\sin^2\theta + \frac{2\sigma^3 - 1}{2\sigma^6} & \sigma > 1.
\end{array}\right.
\end{equation}
\end{subequations}
where the arbitrary pressure constant has been chosen without loss of generality such that $p\to0$ as $\sigma\to\infty$.  Substitution of the solution \eqref{eq:Hsol} into (\ref{eq:gov}$a$--$b$) in the limit $\Ro\to\infty$ shows that Hill's solution satisfies the steady non-rotating Euler equations.  The solution is in the frame of reference of the vortex ring and so the velocity in the far-field tends to $-\hat{\vc{z}}$.  The solution is axisymmetric and swirl-free and the velocity can be represented by a streamfunction, $\psi$, of the form of \eqref{eq:taylorSola}, where
\begin{equation} \label{eq:psiH}
\psi = \left\{
\begin{array}{ll}
\displaystyle \frac{3\sigma^2(1-\sigma^2)}{4}\sin^2\theta & \sigma \leqslant 1 \\[2ex]
\displaystyle  -\frac{(\sigma^3-1)}{2\sigma}\sin^2\theta & \sigma > 1
\end{array}\right.
\end{equation}
The arbitrary constant that may be added to the streamfunction is chosen such that $\psi=0$ on the boundary of the spherical vortex.  The flow is irrotational for $\sigma > 1$ but there is an azimuthal vorticity component that increases linearly in the cylindrical radial direction inside the vortex, $\vc{\omega} = \frac{15}{2}\sigma\sin\theta\,\hat{\vc{\phi}}$ for $\sigma < 1$.  The pressure and velocity fields are continuous across the boundary of the spherical vortex, $\sigma = 1$.

We now make the following observation; if $\vc{u}(\sigma, \theta)$ together with a corresponding pressure field, $p(\sigma, \theta)$, solves the non-rotating Euler equations and $\vc{u}$ can be represented by a streamfunction $\psi(\sigma, \theta)$ in the form of \eqref{eq:taylorSola} with $w=0$, then 
\begin{equation} \label{eq:inner1}
\vc{u} = \left(\frac{1}{\sigma^2\sin\theta}\pd{\psi}{\theta}, -\frac{1}{\sigma\sin\theta}\pd{\psi}{\sigma}, -\frac{\sigma\sin\theta}{2\Ro}\right),
\end{equation}
must solve the rotating Euler equations with the same pressure field $p$.  This is because the swirling component in \eqref{eq:inner1}, $w = -(2\Ro)^{-1}\sigma\sin\theta$, exactly cancels the background rotation of the fluid.  If the flow is viewed from the laboratory frame of reference it is identical to the non-rotating solution.  Thus, we have that \eqref{eq:inner1} with $\psi$ given by Hill's inner non-rotating solution (\eqref{eq:psiH} for $\sigma <1$), and the pressure, $p$ (\eqref{eq:pH} for $\sigma < 1$) automatically satisfies the incompressibility condition and the rotating equations of motion (\ref{eq:gov}$a$--$b$) exactly.  The solution is Hill's non-rotating spherical vortex described in a rotating frame of reference.  To distinguish this solution from the classical solution we shall refer to it as the `swirling' Hill's spherical vortex, even though the azimuthal component of the velocity field exactly cancels the background rotation of the fluid.  The observation that Taylor's towed sphere in his experiments did not precess about the axis of rotation gives rise to the possibility that there may exist a form of Taylor's solution that can be matched onto the swirling Hill's spherical vortex by enforcing a different choice of boundary conditions to \eqref{eq:taylorbcsf} when solving \eqref{eq:taylor8}.

To match a solution of the form \eqref{eq:taylorSol} onto the swirling Hill's spherical vortex such that the velocity field is continuous across the boundary $\sigma = 1$, we require a solution to \eqref{eq:taylor8} that will yield
\begin{subequations}
\begin{equation}
\vc{u}(1, \theta) = \left(0, \frac{3\sin\theta}{2}, -\frac{\sin\theta}{2\Ro}\right),
\qquad
\lim_{\sigma\to\infty} \vc{u}(\sigma,\theta) = \left(-\cos\theta, \sin\theta, 0\right),
\end{equation}
which in terms of $f$ is
\begin{equation} \label{eq:myfbcs}
f(1) = 0, 
\quad
f'(1) = -\frac{3}{2},
\quad
\lim_{\sigma\to\infty} \frac{f(\sigma)}{\sigma^2} = -\frac{1}{2}.
\end{equation}
\end{subequations}
Substitution of these boundary conditions at $\sigma=1$ (assuming $|f''(1)|<\infty$) into \eqref{eq:pressure} shows that if such an $f$ can be found, the pressure on the boundary will be given by $p(1,\theta) = \frac{1}{2} - \frac{9}{8}\sin^2\theta$, exactly matching the pressure on the boundary of the swirling Hill's spherical vortex (see \eqref{eq:pH} at $\sigma = 1$).  The required solution to \eqref{eq:taylor8} satisfying \eqref{eq:myfbcs} is given by
\begin{equation} \label{eq:fsol}
f(\sigma) = -\frac{\sigma^2}{2} + \frac{1}{2\sigma}\left\{\sigma\cos\left(\frac{\sigma - 1}{\Ro}\right) - \Ro\sin\left(\frac{\sigma - 1}{\Ro}\right)\right\}.
\end{equation}
The condition that $|f''(1)|<\infty$ is satisfied as $f''(1) = -(2\Ro^2)^{-1}$ and so the conditions of continuity of velocity and pressure across the boundary are satisfied.  The solution has the property that Hill's classical solution is recovered in the limit $\Ro\to\infty$.  The streamfunction of the solution is given by 
\begin{equation}  \label{eq:sol}
\psi = \left\{
\begin{array}{ll}
\displaystyle \frac{3\sigma^2(1 - \sigma^2)}{4}\sin^2\theta & \sigma \leqslant 1 \\[2ex]
\displaystyle -\frac{\sigma^2\sin^2\theta}{2}\left\{1 -  \frac{1}{\sigma^3}\left[\sigma\cos\left(\frac{\sigma - 1}{\Ro}\right) - \Ro\sin\left(\frac{\sigma - 1}{\Ro}\right)\right]\right\} & \sigma >1
\end{array}
\right.
\end{equation}
Thus, we have a complete steady solution to the nonlinear rotating Euler equations whose inner solution is Hill's spherical vortex with an additional swirling component in the azimuthal direction that cancels out the background rotation of the fluid.  This inner solution matches onto an outer solution, with continuous velocity and pressure across the boundary of the vortex. In the far-field the velocity tends to the free-stream velocity $-\hat{\vc{z}}$ and pressure tends to the hydrostatic pressure field $p = \sigma^2\sin^2\theta/(8\Ro^2)$.  We observe that $\psi$ is even in the Rossby number and so the waves that form in a meridional plane ($\phi = \textrm{const.}$) oscillate according only to the magnitude of the rotation of the system, and not the sign of the direction of rotation, as might be expected on physical grounds.  Similarly, as a result of \eqref{eq:taylorSolb} and \eqref{eq:sol}, the azimuthal velocity is odd in the Rossby number and so the azimuthal flow field is reversed if the sign of the direction of rotation of the system is reversed.  We see from \eqref{eq:sol} that the wavelength of the inertial waves in the outer fluid is $2\pi\Ro$ as in \citet{taylor22}.

\begin{figure} % Image created with vrPaperFig2.m
\begin{center}
%\setlength{\unitlength}{1pt}
%\begin{picture}(350, 227)(0, 0)
%\put(0, 0){\includegraphics[bb = 149   112   573   959, clip = TRUE, width = 110pt]{images/psiField1.eps}}
%\put(-8, -1){{\scriptsize -4}}		\put(-8, 26){{\scriptsize -3}}	\put(-8, 54){{\scriptsize -2}}	\put(-8, 81){{\scriptsize -1}}	\put(-8, 108){{\scriptsize \phantom{-}0}}
%\put(-1, -6){{\scriptsize 0}}		\put(26, -6){{\scriptsize 1}}		\put(53, -6){{\scriptsize 2}}		\put(80, -6){{\scriptsize 3}}		\put(108, -6){{\scriptsize 4}}
%\put(-8, 135){{\scriptsize \phantom{-}1}}	\put(-8, 163){{\scriptsize \phantom{-}2}}	\put(-8, 190){{\scriptsize \phantom{-}3}}	\put(-8, 217){{\scriptsize \phantom{-}4}}
%\put(53, -14){$r$}
%\put(5, 205){$(a)$}	\put(-14, 108){$z$}
%%
%\put(120, 0){\includegraphics[bb = 149   112   573   959, clip = TRUE, width = 110pt]{images/psiField5.eps}}
%\put(119, -6){{\scriptsize 0}}		\put(146, -6){{\scriptsize 1}}		\put(173, -6){{\scriptsize 2}}		\put(200, -6){{\scriptsize 3}}		\put(228, -6){{\scriptsize 4}}
%\put(173, -14){$r$}
%\put(125, 205){$(b)$}
%%
%\put(240, 0){\includegraphics[bb = 149   112   573   959, clip = TRUE, width = 110pt]{images/psiField6.eps}}
%\put(239, -6){{\scriptsize 0}}		\put(266, -6){{\scriptsize 1}}		\put(293, -6){{\scriptsize 2}}		\put(320, -6){{\scriptsize 3}}		\put(348, -6){{\scriptsize 4}}
%\put(293, -14){$r$}
%\put(245, 205){$(c)$}
%\end{picture} \\[3ex]
%
% dvips -o vrPaperFig2.eps -pp6 -O -97pt,-122pt -T 370pt,241pt vrPaper
%\includegraphics[clip = TRUE]{vrPaperFig2.eps}\\[-2ex]
\includegraphics{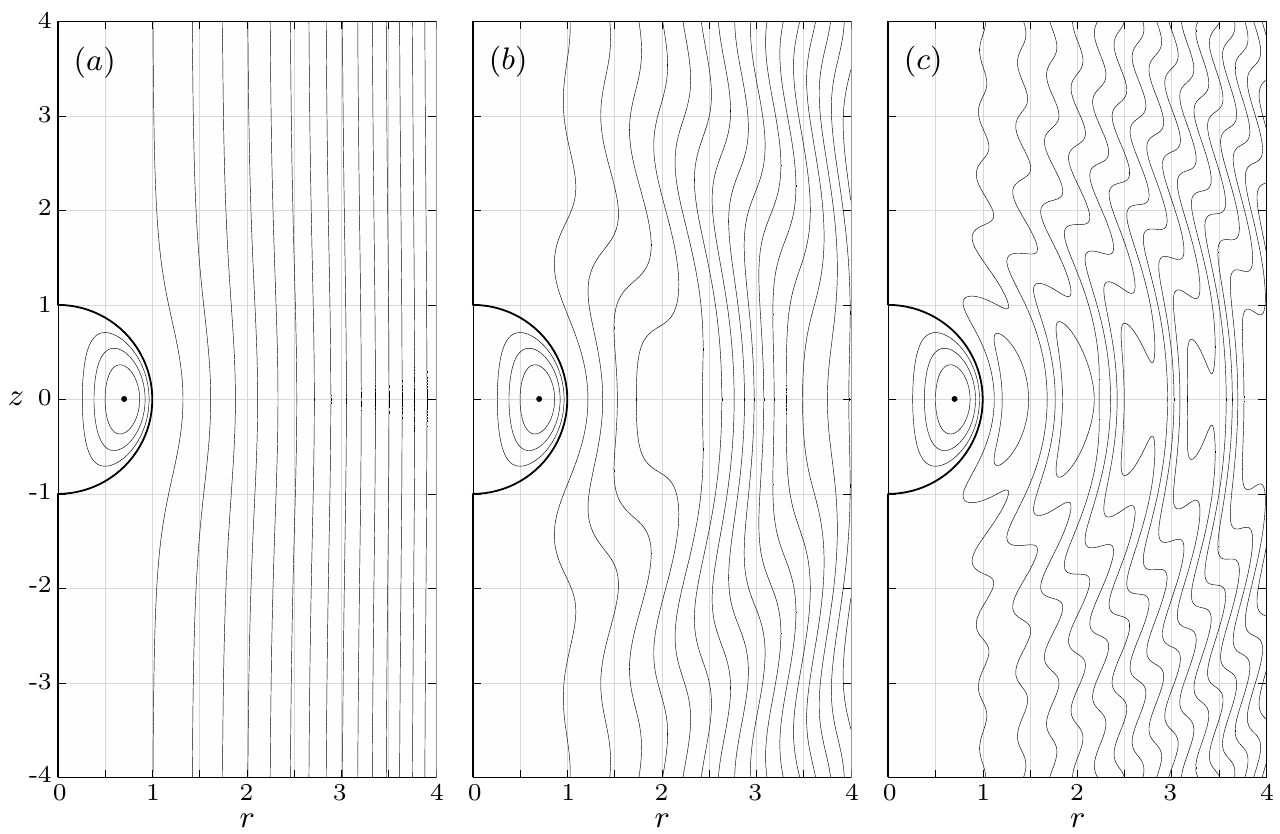}\\[-2ex]
\caption{\label{fig:RotHillSols}Solutions of the streamfunction for $\Ro = \infty$, $\frac{1}{4}$, $\frac{1}{10}$ for $(a)$--$(c)$ respectively.  The meridional stagnation point in the spherical vortex is indicated by a small black circle and is located at $r = 2^{-1/2}$, $z=0$.  In plot $(c)$ the formation of closed streamlines externally to the spherical vortex may be observed, indicating regions of fluid that are transported with the spherical vortex.}
\end{center}
\end{figure}

%****************************************************************************************

\subsection{\label{sec:results}Results}

%****************************************************************************************

Figure \ref{fig:RotHillSols} shows the spherical vortex and the induced flow outside the spherical vortex for three different values of the Rossby number.   The meridional velocity components are always zero at $r = 2^{-1/2}$, $z=0$, indicated by black circles.  Plot $(a)$ is of the non-rotating Hill's spherical vortex that corresponds to the limit $\Ro\to\infty$.   As the rate of rotation is increased, and the Rossby number reduces from $\Ro=\frac{1}{4}$ in plot $(b)$ to $\Ro = \frac{1}{10}$ in plot $(c)$, the onset of inertial waves can be observed in the outer fluid.  Below a critical Rossby number it can be seen that these waves begin to overturn.  We also observe that closed streamlines appear in the outer fluid in plot $(c)$.  These closed streamlines show that above a critical rotation rate, regions of fluid are transported with the spherical vortex, externally to the spherical vortex in the form of concentric vortex rings.

\begin{figure} % Image created with vrPaperFig3.m
\begin{center}
%\resizebox{\textwidth}{!}{
%\setlength{\unitlength}{1pt}
%\begin{picture}(394, 145)(0, 0)
%\put(34, 14){\includegraphics[bb = 72    45   508   390, clip = TRUE, width = 160pt]{images/criticalPoints.eps}}
%\put(112, 0){$\sigma$}		\put(0, 75){$\Ro$}	\put(0, 134){$(a)$}
%\put(18, 13){{\scriptsize 0.05}}	\put(18, 44){{\scriptsize 0.10}}	\put(18, 75){{\scriptsize 0.15}}	\put(18, 107){{\scriptsize 0.20}}	\put(18, 138){{\scriptsize 0.25}}
%\put(32, 7){{\scriptsize 1}}	\put(64, 7){{\scriptsize 2}}	\put(96, 7){{\scriptsize 3}}	\put(128, 7){{\scriptsize 4}}	\put(160, 7){{\scriptsize 5}}	\put(191, 7){{\scriptsize 6}}
%%
%\put(234, 14){\includegraphics[bb = 72    45   508   390, clip = TRUE, width = 160pt]{images/turningPoints.eps}}
%\put(222, 13){{\scriptsize -12}}	\put(222, 33){{\scriptsize -10}}	\put(222, 54){{\scriptsize \phantom{1}-8}}	\put(222, 75){{\scriptsize \phantom{1}-6}}	
%\put(222, 96){{\scriptsize \phantom{1}-4}}	\put(222, 117){{\scriptsize \phantom{1}-2}}	\put(222, 138){{\scriptsize \phantom{1-}0}}
%\put(233, 7){{\scriptsize 1}}	\put(264, 7){{\scriptsize 2}}	\put(296, 7){{\scriptsize 3}}	\put(328, 7){{\scriptsize 4}}	\put(360, 7){{\scriptsize 5}}	\put(391, 7){{\scriptsize 6}}
%\put(214, 75){$\psi$}	\put(312, 0){$\sigma$} \put(210, 134){$(b)$}
%\end{picture}}
%
%dvips -o vrPaperFig3.eps -pp7 -O -97pt,-122pt -T 386pt,140pt vrPaper
%\includegraphics[clip = TRUE]{vrPaperFig3.eps}
\includegraphics{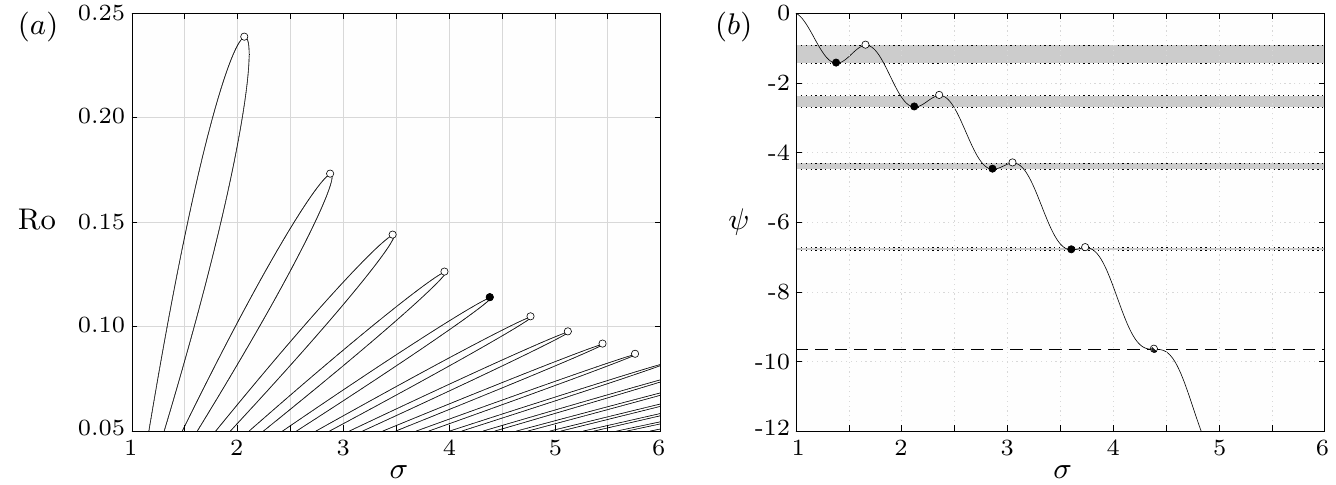}
\caption{\label{fig:critPoints}$(a)$ Zeros of \eqref{eq:wb1}; as the rotation rate increases the Rossby number, $\Ro$, decreases and new branches of the solution are found.  The critical rotation rates at which new branches of solution are found are given by zeros of the system \eqref{eq:wb}, indicated by circles.  The fifth critical Rossby $\Ro = \Ro_c^{(5)} \approx 0.114$ is indicated by the black circle. $(b)$ The streamfunction \eqref{eq:sol} for $\theta = \pi/2$ and $\Ro = \Ro_c^{(5)}$.  This plot corresponds to the $z=0$ transect of figure \ref{fig:psi}.  The black circles that occur at local minima in $\psi$ correspond to black stagnation points  in figure \ref{fig:psi}.  The white circles that occur at local maxima in $\psi$ correspond to the white stagnation points in figure \ref{fig:psi}.  For $\psi$ lying between pairs of black and white turning points, indicated by the grey regions, the function $\sigma(\psi)$ is multi-valued and this corresponds to closed streamlines in the flow.  At the chosen Rossby number the fifth set of closed streamlines is about to appear at the inflection point indicated by the dashed line, corresponding to the cusp at $\sigma \approx 4.39$, $\theta = \pi/2$ (black-white circle) in figure \ref{fig:psi}.}
\end{center}

\end{figure}

The critical rotation rate at which the onset of overturning is observed may be found by considering the turning points of the streamlines.  If a given streamline, $\psi = \textrm{const.}$ for $\sigma>1$, is parameterized by $\theta = \theta(\sigma)$ then we have overturning when $\ud \theta(\sigma)/\ud\sigma = 0$, that is, when
\begin{subequations} \label{eq:wb}
\begin{equation} \label{eq:wb1}
2\Ro\,\sigma^3 + \Ro\,\sigma\cos\left(\frac{\sigma - 1}{\Ro}\right) + \left(\sigma^2 - \Ro^2\right)\sin\left(\frac{\sigma - 1}{\Ro}\right) = 0.
\end{equation}
This expression has a different number of branches of solution for a given Rossby number, $\Ro$, as is shown in figure \ref{fig:critPoints}$a$.  It can be seen that above a critical Rossby number no solutions to \eqref{eq:wb1} exist.  There is therefore a minimum rate of rotation below which no closed streamlines in the outer fluid are formed and no fluid is carried along with the spherical vortex.  The first critical rotation rate occurs when $\ud \Ro(\sigma)/\ud \sigma = 0$ where $\Ro = \Ro(\sigma)$ is determined by \eqref{eq:wb1}.  This condition is given by
\begin{equation} \label{eq:wb2}
6\Ro^2\sigma + \Ro\sin\left(\frac{\sigma - 1}{\Ro}\right) + \sigma\cos\left(\frac{\sigma - 1}{\Ro}\right) = 0.
\end{equation}
\end{subequations}
The critical points determined by simultaneous solutions of \eqref{eq:wb1} and \eqref{eq:wb2} are shown as circles in figure \ref{fig:critPoints}$a$ and we denote the critical Rossby numbers as $\Ro = \Ro_c^{(n)}$ and the corresponding critical radii as $\sigma = \sigma_c^{(n)}$ for $n=1,2,3, \ldots$.  The first critical rotation rate, $\Ro_c^{(1)}$, that represents the minimum rotation rate for which a closed streamline forms in the flow is found numerically to be $\Ro_c^{(1)} \approx 0.239$.  This rotation rate lies between those shown in figure \ref{fig:RotHillSols}$b$ and \ref{fig:RotHillSols}$c$.  The  corresponding radius, $\sigma_c^{(1)}\approx2.07$ and $\theta = \pi/2$ gives the location at which the overturning first occurs.  For a given Rossby number, the associated number of branches of solutions of \eqref{eq:wb1} corresponds to the number of closed streamlines in the flow outside the spherical vortex, and hence corresponds to the number of regions of fluid that are advected with the spherical vortex.

\begin{figure} % Image created with vrPaperFig4.m
\begin{center}
%\setlength{\unitlength}{1pt}
%\begin{picture}(360, 246)(0, 0)
%\put(0,0){\includegraphics[bb = 168   196  1184   875, clip = TRUE, width = 360pt]{images/siblingRings.eps}}
%\put(-1, -6){{\scriptsize 0}}		\put(59, -6){{\scriptsize 1}}		\put(119, -6){{\scriptsize 2}}	\put(179, -6){{\scriptsize 3}}
%\put(238, -6){{\scriptsize 4}}	\put(298, -6){{\scriptsize 5}}	\put(358, -6){{\scriptsize 6}}
%\put(179, -14){$r$}
%\put(-8, -1){{\scriptsize -2}}		\put(-8, 59){{\scriptsize -1}}	\put(-8, 119){{\scriptsize \phantom{-}0}}
%\put(-8, 178){{\scriptsize \phantom{-}1}}	\put(-8, 238){{\scriptsize \phantom{-}2}}
%\put(-14, 119){$z$}
%\end{picture} \\[3ex]
%
%dvips -o vrPaperFig4.eps -pp8 -O -94pt,-122pt -T 378pt,260pt vrPaper
%\includegraphics[clip = TRUE]{vrPaperFig4.eps}
\includegraphics{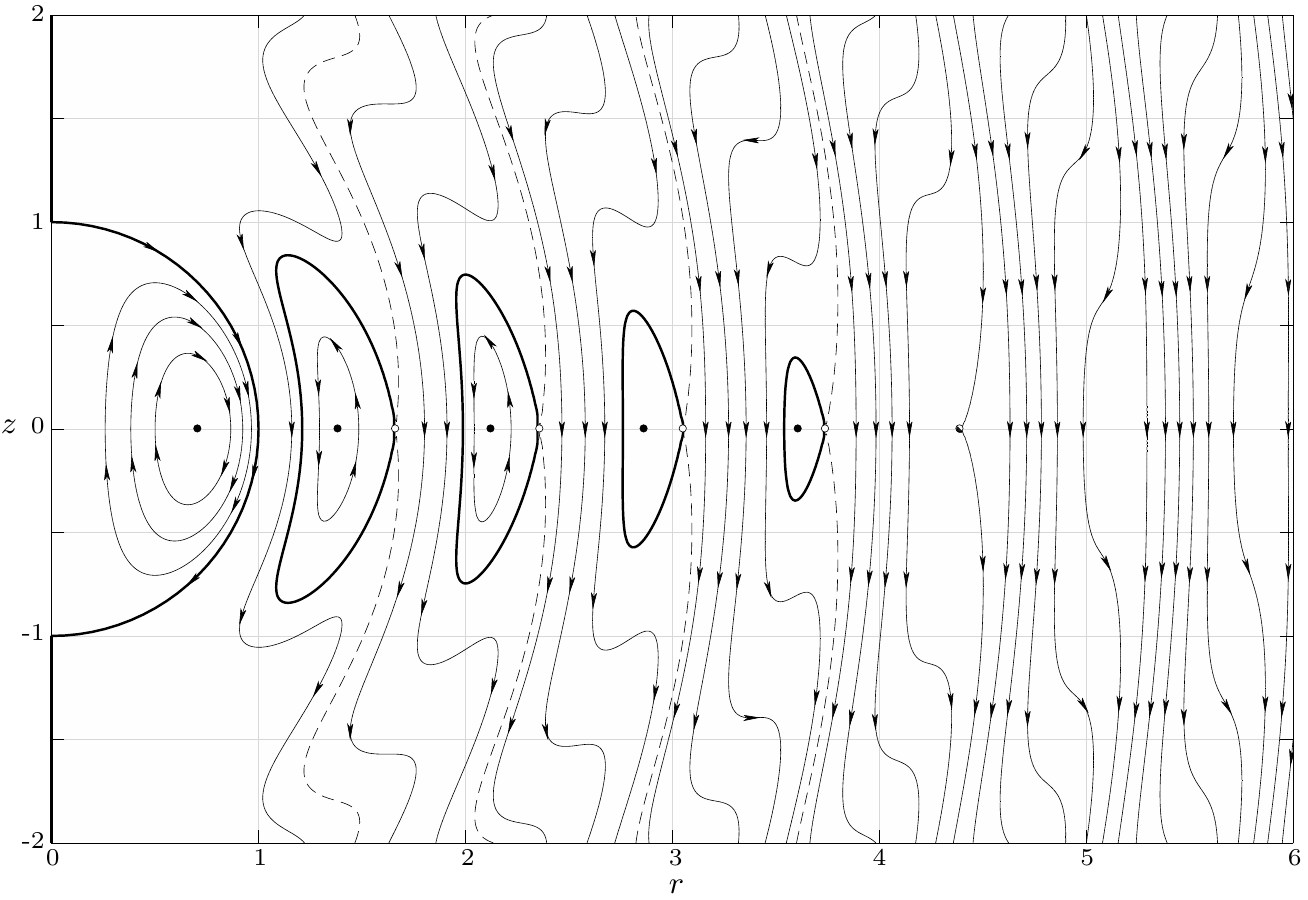}
\caption{\label{fig:psi}A meridional slice through the flow field in the frame of reference moving with the vortex ring for $\Ro = \Ro_c^{(5)} \approx 0.114$. This is the fifth critical Rossby number, we can see four sibling vortex rings (bold lines) have been created and there is a cusp that has formed (black-white circle at $\sigma\approx4.39$) in a streamline that will form the next sibling vortex ring when $\Ro < \Ro_c^{(5)}$.}
\end{center}
\end{figure}

The value of the streamfunction on the closed streamlines and the location of stagnation points in the flow can be found by considering $\psi$ on $\theta = \pi/2$.  Figure \ref{fig:critPoints}$b$ shows $\psi(\sigma, \pi/2)$ for $\sigma>1$ and $\Ro = \Ro_c^{(5)} \approx 0.114$, the fifth critical rotation rate (indicated by the black circle in figure \ref{fig:critPoints}$a$).  The turning points in $\psi$ can be seen to appear in pairs of local minima (black circles) and local maxima (white circles).  For values of $\psi$ between the local minima and maxima, indicated by the grey bands, the function $\sigma(\psi)$ is multi-valued and this corresponds to closed streamlines in the flow.  As the flow considered is exactly at the fifth critical rotation rate, the fifth pair of local minima and maxima coincide at the inflection in $\psi$ where $\sigma\approx4.39$ and $\psi\approx-9.65$.

Figure \ref{fig:psi} shows streamlines of the flow induced by a Hill's spherical vortex in a rotating fluid at $\Ro = \Ro_c^{(5)}$ corresponding to the rotation rate in figure \ref{fig:critPoints}$b$.  The stagnation point in the spherical vortex is, as in the classical solution, at $\sigma = 2^{-1/2}$, $\theta = \pi/2$.  In the outer flow there can be seen to be four closed streamlines, indicated in bold.  We observe that the direction of advection around these closed streamlines (anti-clockwise) is opposite to that in the spherical vortex (clockwise), though the vorticity in the closed streamlines may change sign.  The black stagnation points in the closed streamlines correspond to the black local minima in figure \ref{fig:critPoints}$b$.  The white stagnation points, on the boundary of the closed streamlines, correspond to the local maxima in \ref{fig:critPoints}$b$.  A fifth `closed streamline' is about to form at the cusp indicated by the black-white circle at $\sigma\approx4.39$.  This corresponds to the inflection in $\psi$ in figure \ref{fig:critPoints}$b$.

%****************************************************************************************

\section{\label{sec:conc}Conclusions}

%****************************************************************************************

Following Taylor's (\citeyear{taylor22}) observation that the streamlines formed by a steadily towed, vanishingly small, sphere along the axis of rotation of a rotating fluid resemble those of a Hill's spherical vortex \citep{hill}, we have derived an explicit solution of the rotating Euler equations that supports a steadily propagating spherical vortex.  The spherical vortex swirls in such a way as to exactly cancel the background rotation of the fluid, in that way mimicking the behaviour of the sphere towed through the rotating tank in Taylor's experiments, and in the manner of experimental observations in developed vortex rings \citep[see e.g.,][]{eisenga, verziccoEtAl}.  Thus the spherical vortex is exactly Hill's spherical vortex, but cast in a rotating frame of reference.  This inner solution matches onto an outer solution that exhibits nonlinear inertial waves.  As the rotation rate is increased the amplitude of the waves is observed to grow until, at a critical rotation rate of $\Ro \approx 0.239$, wave overturning may be observed and closed streamlines form in the outer fluid.  The closed streamlines represent vortex rings that are concentric to the spherical vortex on the axis and which propagate with the spherical vortex, but rotate with the opposite sense.  As the rotation rate is increased beyond subsequent critical Rossby numbers, new `sibling' vortex rings are added to the vortex ring family propagating along the axis of rotation.

%****************************************************************************************

%\bibliographystyle{plainnat}
%\bibliography{vrPaper}

\end{document}